\newif\ifPDF
\documentclass[usenatbib]{mn2e}
\ifPDF
  \usepackage[T1]{fontenc}
  \usepackage{times}
  \usepackage[hypertex,colorlinks=true,citecolor=blue]{hyperref}
  \usepackage{aeguill}
  \usepackage[pdftex]{graphicx,color}
  \usepackage{subfigure}
  \usepackage{afterpage}
\else
  \usepackage[T1]{fontenc}
  \usepackage{times}
  \usepackage[dvips]{graphicx,color}
  \usepackage{subfigure}
  \usepackage{afterpage}
\fi
\title[Determination of the orbital parameters of binary pulsars]
{Determination of the orbital parameters of binary pulsars}
\author[Bhaswati Bhattacharyya, Rajaram Nityananda]
{Bhaswati Bhattacharyya,$^1$
 Rajaram Nityananda$^1$\\\\
 $^1$National Centre for Radio Astrophysics, TIFR, Pune University Campus, Post Bag 3,
Pune 411 007, India\\}

\date{Accepted. Received}
\begin{document}
\label{firstpage}
\maketitle
\pagerange{\pageref{firstpage}--\pageref{lastpage}} \pubyear{2007}
\def\LaTeX{L\kern-.36em\raise.3ex\hbox{a}\kern-.15em
    T\kern-.1667em\lower.7ex\hbox{E}\kern-.125emX}
\begin{abstract}
We present a simple method for determination of the orbital 
parameters of binary pulsars, using  data on the pulsar 
period at multiple observing epochs. This method uses the 
circular nature of the velocity space orbit of Keplerian 
motion and produces preliminary values based on two one 
dimensional searches. Preliminary orbital parameter values 
are then refined using a computationally efficient linear 
least square fit. This method works for random and sparse 
sampling of the binary orbit. We demonstrate the technique 
on (a) the highly eccentric binary pulsar PSR J0514$-$4002 
(the first known pulsar in the globular cluster NGC 1851) 
and (b) 47 Tuc T, a binary pulsar with a nearly circular 
orbit.   
\end{abstract}
\section{Introduction}
                                                 \label{sec:intro}       
Knowledge of the orbital parameters of binary pulsars is 
necessary for coherent timing  and for investigation of 
different properties of the pulsar and the companion star. 
Determination of the orbital parameters is important for 
newly discovered pulsars, to plan follow up observations 
at different epochs.

With the movement of the binary pulsar in its orbit around 
the center of mass, the projected velocity of the pulsar in 
the line of sight direction ($v_{l}$) changes and as a 
consequence the observed pulsar period ($P_{obs}$) changes. 
The modulation in $v_{l}$ (i.e. in ${P_{obs}}$) is governed 
by the orbital parameters of the binary system. So it is 
possible to get information about the orbit by studying the 
evolution of ${P_{obs}}$. Five orbital parameters, namely, 
the binary orbital period ($P_b$), orbital eccentricity ($e$), 
projection of the semi major axis on the line of sight 
($a_1 sin\:i$, $i$ being the angle between the orbit and the 
sky plane), longitude of periastron ($\omega$) and the epoch 
of periastron passage ($T_o$) can be determined from radial 
velocity/observed pulsar period data (in the Newtonian, i.e 
non-relativistic regime). These orbital parameters of binary 
pulsar systems can be determined by fitting a Keplerian model 
to the pulsar period versus epoch of observation data. The 
usual methods require simultaneous fit to many parameters 
and need an initial guess. Such methods need dense sampling 
of period measurements at different epochs during the pulsar 
orbital period. Overcoming some of these factors, 
\cite{Freire_etal_b} proposed a new method for determination 
of the orbital parameters of binary pulsars. They utilised 
information on periods and period derivatives at multiple 
observing epochs of the kind used in surveys, and extracted
orbital parameter values. They successfully determined the 
orbital parameters of binary pulsars with nearly circular 
orbits.

This work presents an alternative approach to orbital 
parameter determination using the observing epoch versus 
pulsar period data, without requiring information about 
pulsar period derivatives. We demonstrate the method by 
estimating the orbital parameters of the binary pulsar 
PSR J0514$-$4002A, the first known pulsar in the globular 
cluster NGC 1851 \citep{Freire_etal_04}, and PSR J0024$-$7204, 
a binary pulsar in globular cluster 47 Tucanae, referred to 
as 47 Tuc T hereafter \citep{Camilo_etal}. In 
Sect.\ref{sec:det_orbital_P} we describe the method for 
preliminary determination of the orbital parameters. 
Sect.\ref{sec:det_orbital_para} presents a method for 
refinement of the determined orbital parameters. In 
Sect.\ref{sec:discussion} we compare the orbital parameters 
determined in this work with those available in the 
literature and discuss the advantages of our method.

\section{Preliminary determination of orbital parameters}                  \label{sec:det_orbital_P} 
\subsection{Binary orbital period ($P_b$)}
The observed pulsar period ($P_{obs}$) versus epoch of 
observation data set is folded with wide range of trial 
orbital periods ($P_{b}$). Corresponding to each trial 
value of $P_{b}$, we get, $P_{obs}$ versus orbital phase 
($\phi=2\pi t /P_b$, $t$ being the time measured from the 
periastron). For every set of folded data we calculate a 
parameter $-$ roughness ($R$) $-$ which we define as the 
summation of squared differences of $P_{obs}$ between the 
adjacent pairs of $\phi$. Therefore,
\begin{equation}  
R=\sum_{i=1}^{n} ({{P_{obs}}(i)-{P_{obs}}(i+1)})^2 
\label{eqn1}
\end{equation}
where $n$ represents the total number of data points. These 
points are sorted in order of orbital phase, which will be 
different for different choices of trial $P_{b}$. For the optimal 
choice of the trial folding period $P_{b}$, the plot of $P_{obs}$ 
versus $\phi$ is expected to be the smoothest and hence the 
corresponding roughness parameter ($R$) will be minimum. In the 
search of $P_{b}$ the increment ($\Delta P_b$) must be chosen to 
cause small orbital phase shift (i.e. $\Delta \omega_b \: T<<1$) 
over the full data length $T$ (i.e. $(2\pi/P_{b}^2)\Delta P_{b} T<<1$).

As a crosscheck, we apply this method on a simulated 
Keplerian orbit. First, we simulate sparsely and randomly 
sampled epoch of observation versus radial velocity data 
points with a set of arbitrarily chosen $P_{b}$, $e$, 
$\omega$ and $T_o$ values (refer to Eqn. \ref{eqn6} of 
Sect. \ref{sec:det_othorbital_P} for details). Using this 
kind of randomly generated radial velocity data, spanning 
over widely separated epochs, as input we apply the 
smoothness criterion described in Eqn. \ref{eqn1} and the 
true binary orbital period is recovered. There are few 
local minimas where $R$ is lower than the adjacent values 
but there is no comparable minimum as to the strongest 
minimum corresponding to true $P_{b}$. The method 
worked for Keplerian orbits generated with various sets 
of $P_{b}$, $e$, $\omega$ and $T_o$ values, and we could 
reproduce the true periodicity. Hence, to obtain a unique 
solution for $P_{b}$, one need to search for $P_{b}$ within 
a wide range which includes the actual $P_{b}$ with small 
enough step size determined by the criterion 
$(2\pi/P_{b}^2)\Delta P_{b} T<<1$.

\begin{figure}   
\begin{center}
\includegraphics[angle=-90, width=0.47\textwidth]{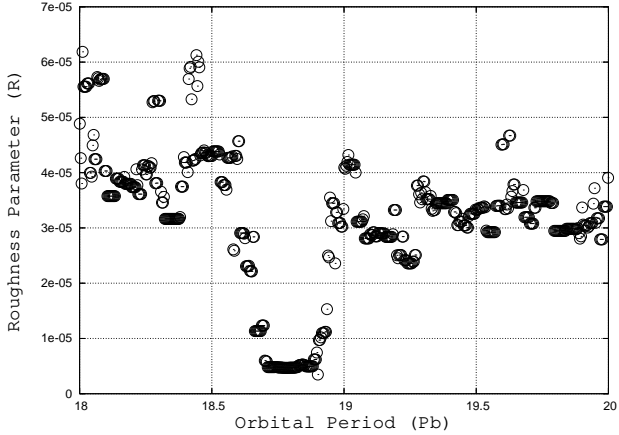}
\caption{Roughness parameter ($R$) plotted against orbital period ($P_b$) for PSR J0514$-$4002A (a zoomed region near minimum $R$)}
\label{fig1}
\end{center}
\end{figure}

\begin{figure}
\begin{center}
\includegraphics[angle=-90, width=0.47\textwidth]{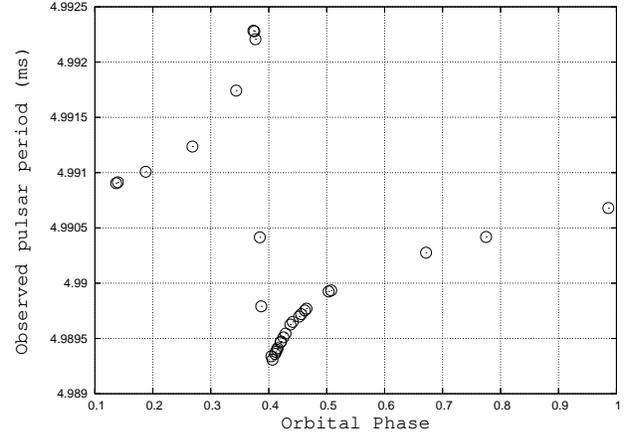}
\caption{Orbital phase ($\phi$) versus observed pulsar period ($P_{obs}$) of PSR J0514$-$4002A after folding the data with $P_b$=18.791 days}
\label{fig2}
\end{center}
\end{figure}

For preliminary determination of $P_{b}$ of PSR J0514$-$40, 
we used $P_{obs}$ versus epoch of observation data from the 
GMRT observations. We used 31 such data points, collected 
over six months, which are similar to the data used for 
\cite{Freire_etal_04}. For the known binary pulsars in 
globular clusters the orbital periods lie in the range 
$P_{b}$ $\sim$ few hours to 256 days (refer to Table 1.1 
of \cite {thesis_Freire}). Initially we try $P_{b}$ starting 
from few hours and up to $300$ days with step size satisfying 
the criterion $(2\pi/P_{b}^2)\Delta P_{b} T<<1$, and determine 
$R$ using Eqn.\ref{eqn1}. Then we narrowed down our search
of the $P_{b}$ around the lowest $R$. Though there are few 
local minima where $R$ is lower than the adjacent values, we 
observe the strongest and rather flat minimum for a range of 
nearby values of ${P_{b}}$ s around 18.79 days, no comparable 
minimum is observed in the range from few hours to 300 days. 
Fig. \ref{fig1} presents the plot of the trial $P_{b}$ 
against the corresponding $R$, zoomed into a region where $R$ 
is minimum. For ${P_{b}}$=18.791 days $R$ is minimum. We 
fold the data with ${P_{b}}$=18.791 days to generate 
$P_{obs}$ versus $\phi$ data set (see Fig.\ref{fig2}).

For the determination of orbital period of 47 Tuc T we 
utilised the 9 data points (provided in \cite{Freire_etal_b}) 
of $P_{obs}$ versus epoch of observation. We determine 
${P_{b}}$=1.1 days which is close to the value estimated 
by \cite{Freire_etal_b}.  
\begin{figure*}
\begin{center}
\includegraphics[angle=0, width=0.93\textwidth]{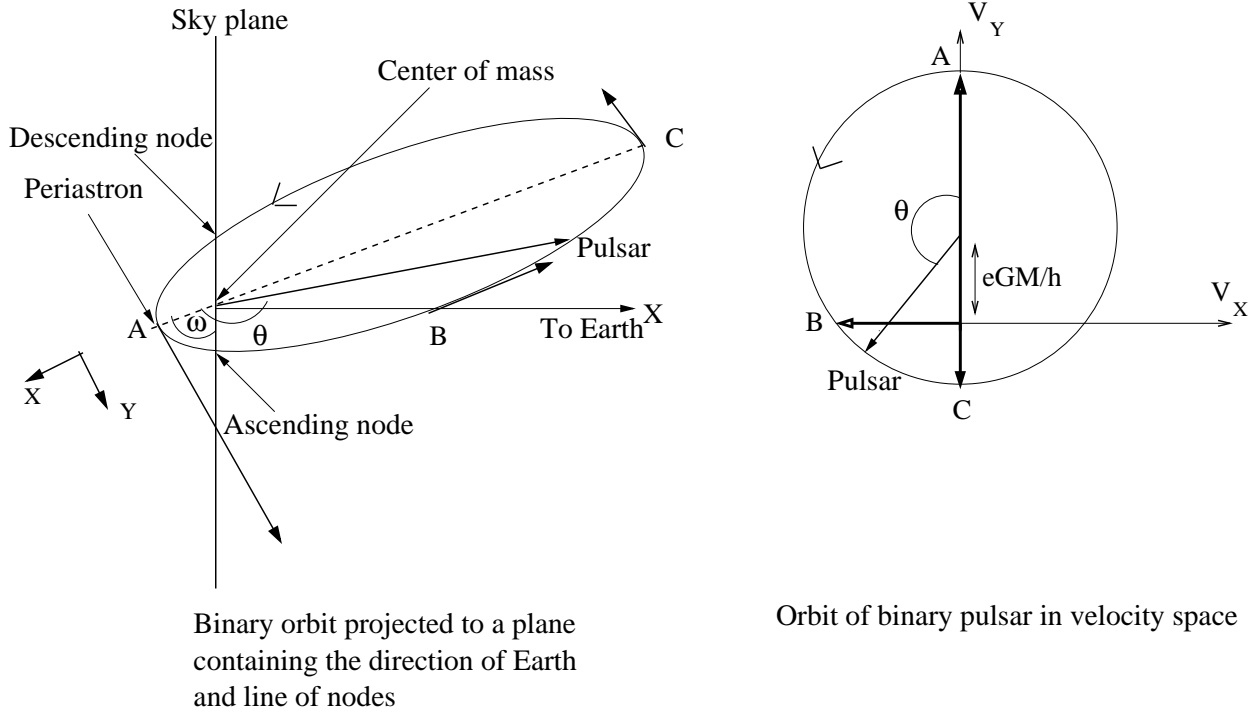}
\caption{Binary orbit}
\label{fig3}
\end{center}
\end{figure*}

\subsection{Other orbital parameters from the hodograph} \label{sec:det_othorbital_P}
The left panel of Fig. \ref{fig3} shows the orbit of a 
binary pulsar around the center of mass of the system, 
projected in a plane containing the direction of Earth and 
the line of nodes (line of intersection of orbital plane 
and the sky plane). 'A' denotes the periastron position 
and $\theta$ is the angle of the pulsar to the periastron, 
also known as 'true anomaly'. 'B' and 'C' are two other 
points in the binary orbit. A rather geometric picture of 
the Kepler's laws using the idea of velocity space is due 
to \cite{Hamilton}. It is not often used and hence described 
briefly below. According to Newton's laws for the path of 
the vector  $(\vec{r}_{pulsar}(t)-\vec{r}_{companion}(t))$ 
(i.e. for the relative orbit of the pulsar with respect to 
the companion star), the relative velocity,
\begin{equation}  
        \Delta \mathbf{v}=-\left(\frac{GM}{r^2}\right) {\Delta t} \;\mathbf{\hat{r}}
\label{eqn2}
\end{equation}
where $G$ is the Gravitational constant and $M$ is the total 
mass of the pulsar and the companion star.\\
From the conservation of angular momentum,
\begin{equation}  
        \Delta \theta = \frac{h}{r^2}\:{\Delta t}
\label{eqn3}
\end{equation}
where $h$ is angular momentum per unit mass.\\
Dividing the absolute value of Eqn. \ref{eqn2} by Eqn. 
\ref{eqn3} we get,
\begin{equation}      
\frac{|\Delta \mathbf{v}|}{\Delta \theta} = \left(\frac{GM}{h}\right)=constant
\label{eqn4}
\end{equation}
The path followed by the velocity vector of a particle is 
called the hodograph. $\Delta\mathbf{v}$ is the arc length 
and $\Delta\theta$ is the angle traversed by the pulsar in 
velocity space. The ratio $(|\Delta\mathbf{v}|/\Delta \theta)$ 
is the radius of curvature of the hodograph. Since the 
radius of curvature is constant, the hodograph is a circle 
for Keplerian motion. The right panel of Fig.\ref{fig3} 
shows the corresponding hodograph of the elliptical binary 
orbit that is shown in the left panel. The center of the 
circle is offset from the origin by $(eGM/h)$ and the 
radius of the circle is $(GM/h)$. 

For a particular eccentricity ($e$) and longitude of 
periastron ($\omega$), the $x$ and $y$ component of velocity 
are given by,
\begin{equation}  
         {v_{x}}=-\frac{GM}{h}sin \: \theta;  \; {v_{y}}=\frac{GM}{h}(cos \: \theta+e) 
\label{eqn5}
\end{equation}
Hence, the relative radial velocity along the projection 
of the line of sight into the orbital plane is given by,
\begin{eqnarray}  
{v_{r}}&=& \left(v_{x}\;cos\:(\pi/2-\omega)+v_{y}\;sin\:(\pi/2-\omega)\right)\nonumber \\
       &=& \left(\frac{GM}{h}\right)\left(sin\:\theta\; sin\:\omega+(cos\:\theta+e)\;cos\:\omega\right) \nonumber\\
       &=&\left(\frac{GM}{h}\right){v_{r}}_{s}
\label{eqn6}
\end{eqnarray}
For $\omega=90\degr$ the observed velocity will be 
antisymmetric (odd) as a function of $\theta$ or time 
measured from periastron. Similarly, for $\omega=0\degr$ 
the observed velocity will be symmetric (even). For other 
intermediate values of $\omega$ the observed velocity will 
be a combination of antisymmetric and symmetric parts in 
the ratio of $sin\:\omega/cos\:\omega$. Plot of the 
antisymmetric versus the symmetric part will be an ellipse 
and the parameters of the ellipse will provide preliminary 
values of the orbital parameters. 

\begin{figure}
\begin{center}
\includegraphics[angle=-90, width=0.47\textwidth]{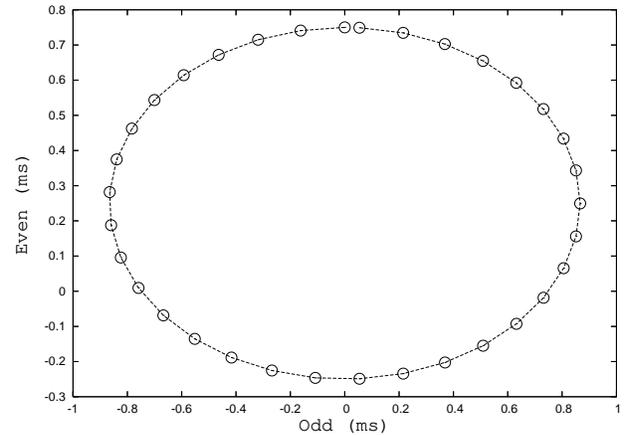}
\caption{${{v_{r}}_{s}^{odd}}$ versus ${{v_{r}}_{s}^{even}}$ (generated for simulated Keplerian orbit with $e=0.5, \omega=60$\degr) and the fitted ellipse for correct choice of $T_o$}
\label{fig4}
\end{center}
\end{figure}

As a crosscheck, we apply this method on simulated 
Keplerian orbits. We simulate ${v_{r}}_{s}$ for trial value 
of $e$, $\omega$ and $T_o$. Corresponding to each ${v_{r}}_{s}$ 
value at a particular orbital phase ($\phi$), we determine 
the ${v_{r}}_{s}$ at conjugate phase ($2\pi-\phi$), using 
Lagrange's interpolation method with three points. The even 
and odd parts are defined as follows,
\begin{equation}      
{v_{r}}_{s}^{even} = ({v_{r}}_{s}(\phi)+{v_{r}}_{s}(2\pi-\phi))/2
\label{eqn7}
\end{equation}
\begin{equation}      
{v_{r}}_{s}^{odd} = ({v_{r}}_{s}(\phi)-{v_{r}}_{s}(2\pi-\phi))/2
\label{eqn8}
\end{equation}
Plot of ${v_{r}}_{s}^{odd}$ versus ${v_{r}}_{s}^{even}$ 
should be an ellipse, for correct choice of $T_o$ 
(Fig.\ref{fig4}). The ratio of major and the minor axes 
of the ellipse gives, $tan\:\omega$, and the shift of the 
origin of the ellipse gives $e$. Using the method 
illustrated in Appendix \ref{sec:appendixA}, we fit an 
ellipse to the ${v_{r}}_{s}^{odd}$ versus 
${v_{r}}_{s}^{even}$ data. $\omega$ and $e$ are recovered 
from the parameters of the best fit ellipse.

\begin{figure*}
\begin{center}
\hbox{
  \includegraphics[angle=-90, width=0.47\textwidth]{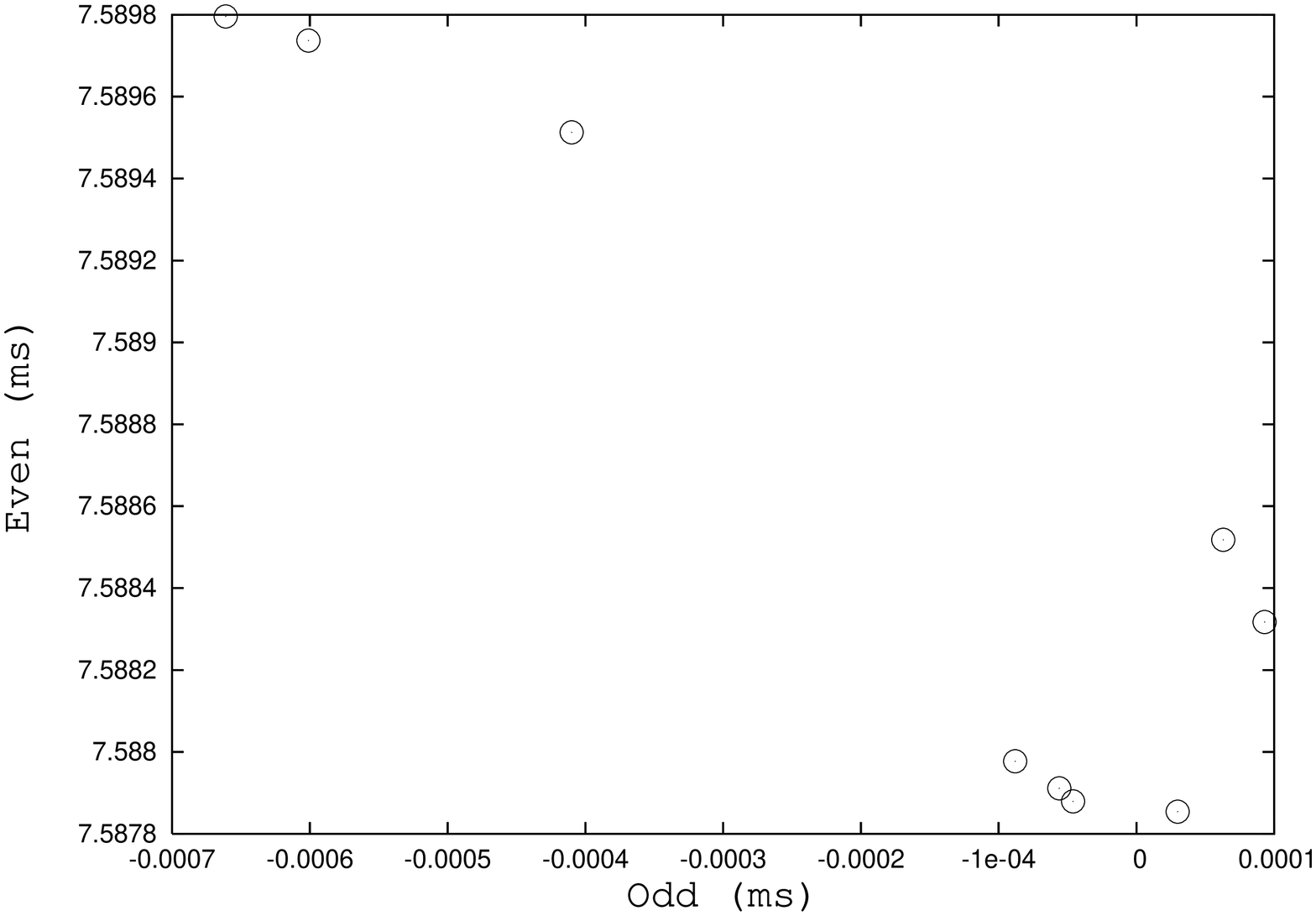}
  \includegraphics[angle=-90, width=0.47\textwidth]{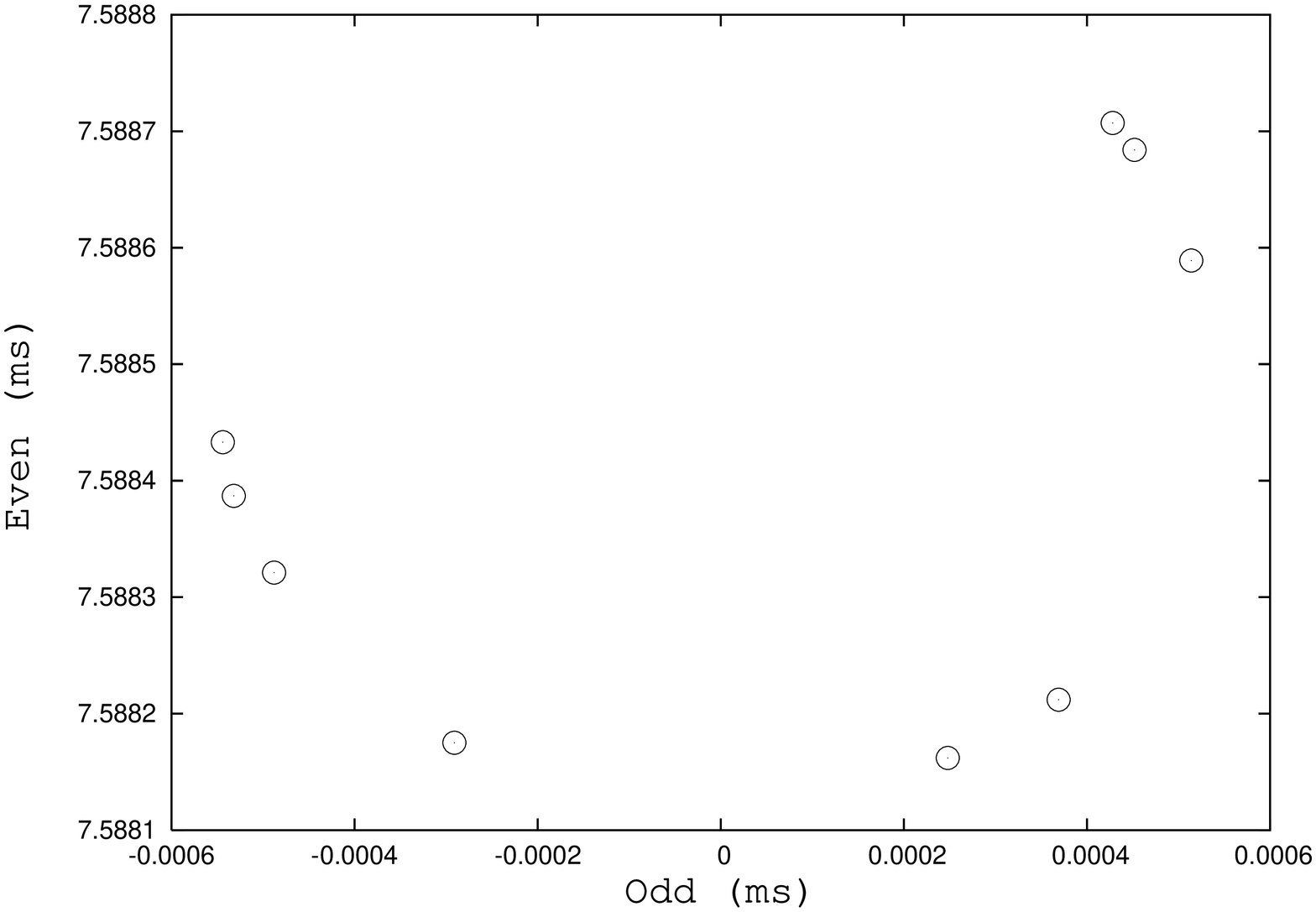}
}
\caption{$P_{obs}^{odd}$ versus $P_{obs}^{even}$ for 47 Tuc T for arbitrary choice of $T_o$ (left panel) and correct choice of $T_o$ with minimum ${\mathbf{\chi ^2}}$ (right panel)}
\label{fig5}
\end{center}
\end{figure*}

Since ${v_{r}}_{s}$ and the observed pulsar period 
($P_{obs}$) will have similar modulations, we construct 
antisymmetric and symmetric parts from the $P_{obs}$. 
Corresponding to each $P_{obs}$ value at a particular 
orbital phase ($\phi$), we determine the $P_{obs}$ at 
conjugate phase ($2\pi-\phi$) using Lagrange's 
interpolation method with three points. The even and 
the odd parts are defined as follows,
\begin{equation}      
P_{obs}^{even} = (P_{obs}(\phi)+P_{obs}(2\pi-\phi))/2
\label{eqn9}
\end{equation}
\begin{equation}      
P_{obs}^{odd} = (P_{obs}(\phi)-P_{obs}(2\pi-\phi))/2
\label{eqn10}
\end{equation}
The plot of $P_{obs}^{odd}$ versus $P_{obs}^{even}$ should 
be an ellipse for correct choice of the periastron passage 
($T_o$). We vary $T_o$, corresponding $P_{obs}^{odd}$ versus 
$P_{obs}^{even}$ are generated, and fit an ellipse to the 
$P_{obs}^{odd}$ versus $P_{obs}^{even}$ plot (Appendix 
\ref{sec:appendixA}). The left panel of Fig. \ref{fig5} 
is the plot of $P_{obs}^{odd}$ versus $P_{obs}^{even}$ for 
real data of 47 tuc T with arbitrary choice of $T_o$. The 
right panel of Fig. \ref{fig5} is the plot of $P_{obs}^{odd}$ 
versus $P_{obs}^{even}$ for real data of 47 tuc T with optimal 
choice of $T_o$ ($T_o$ for which ${\mathbf{\chi ^2}}$ is minimum 
after ellipse fitting). Preliminary values of $e$, $\omega$ are 
obtained from the parameters of the best fit ellipse 
(Appendix \ref{sec:appendixA}).
 
\section {Refinement of the determined orbital parameters}                \label{sec:det_orbital_para}
In this section we take the preliminary determined orbital 
parameters as the initial guess in a linear least squares fit. 
This is now computationally efficient since only a small range 
of the parameters, near the first guess values, has to be 
searched. $P_{obs}$ is determined by the relation,
\begin{equation}   
{P_{obs}}={P_{o}}\;(1+\frac{v_{l}}{c})
\label{eqn11}
\end{equation}
where ${P_{o}}$ is the rest frame period of the binary 
pulsar, $v_{l}$ is the projected velocity of the pulsar in the 
line of sight direction and $c$ is the velocity of light. This 
relationship is valid provided $v_{l}$ is small compared to $c$. 

Following are the steps for determination of orbital parameters:

1. We simulate orbital phase ($\phi$) versus scaled radial 
velocity (${v_{r}}_{s}$) with trial values $P_{b}$, $e$, 
$\omega$, $T_o$ (using Eqn. \ref{eqn6}).

2. To compare the simulated data with the observations we 
need to find out the simulated ${v_{r}}_{s}$ at those 
orbital phase points for which ${P_{obs}}$ is available. 
${v_{r}}_{s}$ at observed orbital phases is obtained by 
using Lagrange's interpolation method with three points.

3. Next we fit a straight line to ${P_{obs}}$ versus 
${v_{r}}_{s}$ and calculate ${\chi ^2}$. 

We repeat this procedure for all the trial combinations of 
orbital parameters. As shown in the Appendix 
\ref{sec:appendixB}, for the right choice of the orbital 
parameters, the plot of ${P_{obs}}$ versus ${v_{r}}_{s}$ 
will be a straight line (see Eqn. \ref{eqn21}). Hence, 
the set of orbital parameters, $P_{b}$, $e$, $\omega$, 
$T_o$, for which the straight line fit is best, i.e. 
${\chi ^2}$ value is minimum, will correspond to the 
optimal choice of orbital parameters. ${\chi ^2}$ 
is minimised so that the expected value for N independent 
data points is N. A change of 1 then corresponds to a 
68\% confidence limit (page 694, \cite {Press_etal}). 
Given the above criterion for change in ${\chi^2}$, the 
optimal grid for any parameter (keeping all the other 
parameter fixed) would have about three points in an 
interval over which the minimum ${\chi ^2}$ (${\chi^2}_{min}$) 
increases by 1 $\sigma$ ($\sigma \sim {\chi^2}_{min}$/N). 
This is the criterion that decide the step size used for 
different trial combinations of the orbital parameters. 
The search for each orbital parameter was continued till 
the ${\chi^2}$ becomes about 1000 $\sigma$ on each side 
of the minima, keeping all the other parameters fixed. It 
is possible to use this method to determine the orbital 
parameters, with out assuming the preliminary values. But 
in that case one has to search a wide range for each of 
the orbital parameters which would be computationally 
expensive. 
The intercept of the fitted straight line will give the 
value of $P_{o}$. Substituting the values of $P_{b}$, $e$, 
$P_{o}$ and the slope of the fitted straight line $S_{fit}$, 
in Eqn. \ref{eqn23}, we can determine the projected semi 
major axis in light seconds, ${a_1 sin(i)}/{c}$. 

{\bf Implementation of the method}\\
{\bf (1) J0514$-$4002 :}
\begin{figure}
\begin{center}
\includegraphics[angle=-90, width=0.47\textwidth]{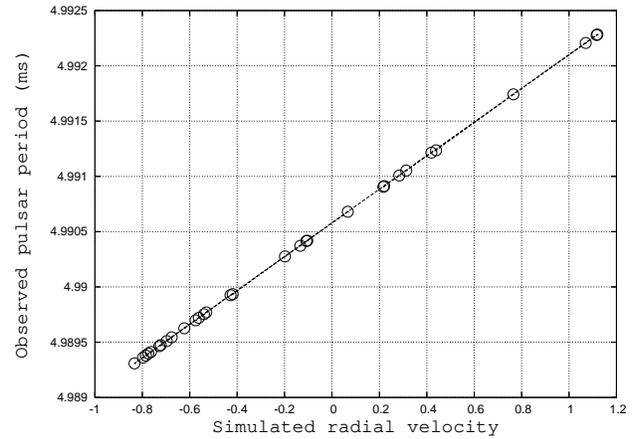}
\caption{Simulated radial velocity (${v_{r}}_{s}$) interpolated at each observing epoch is plotted against observed pulsar period (${P_{obs}}$) for PSR J0514$-$4002}
\label{fig6}
\end{center}
\end{figure}
Fig.\ref{fig6} presents the plot of ${P_{obs}}$ versus 
${v_{r}}_{s}$ (generated with the optimal choice of orbital 
parameters) and the corresponding straight line fit. The 
residual from the best fit straight line are small for all 
the measurements, indicating successful fitting and orbital 
parameter determination. Table. \ref{table:orbital_par1} 
lists the determined orbital parameter values of PSR 
J0514$-$4002. The step size used for the different sets of 
trial of orbital parameters, $P_{b}$, $e$, $\omega$ 
and $T_o$, are also listed in Table. \ref{table:orbital_par1}. 
The uncertainty on the values of each of the orbital parameters 
are calculated from the change of orbital parameter values 
required for $1\sigma$ change in the ${\chi ^2}$ value, keeping 
all the other parameters fixed. The uncertainty quoted in the 
bracket is on the last significant digit of the concerned parameter.\\
{\bf (2) 47 Tuc T :}
\begin{figure}
\begin{center}
\includegraphics[angle=-90, width=0.47\textwidth]{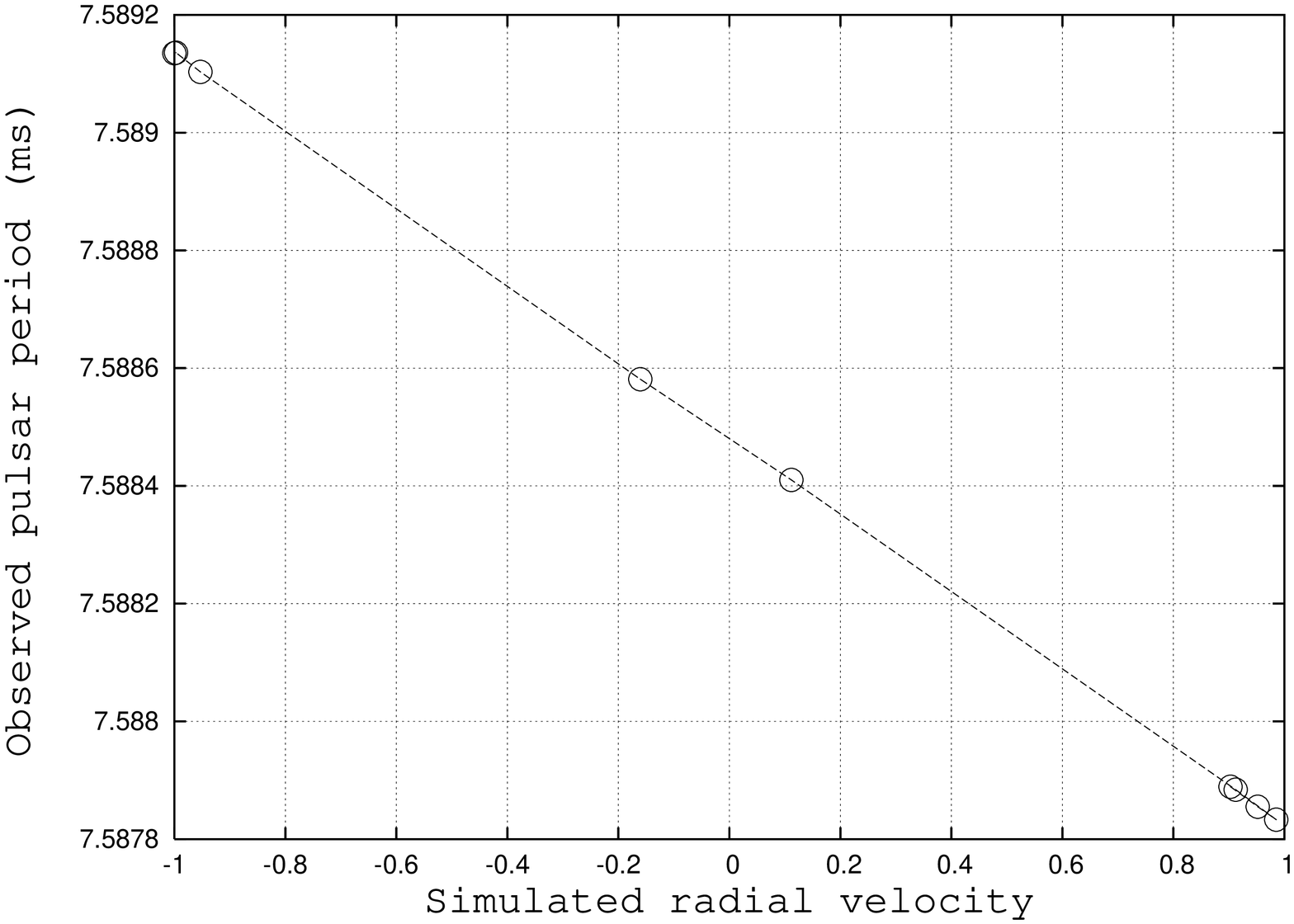}
\caption{Same as Fig. \ref{fig4} for 47 Tuc T}
\label{fig7}
\end{center}
\end{figure}
Fig.\ref{fig7} plots the ${P_{obs}}$ versus the optimal ${v_{r}}_{s}$. 
It is evident that the observational data is well reproduced. 
Determined orbital parameter values and the associated errors
 are listed in Table. \ref{table:orbital_par2}.
\begin{table*}
\begin{minipage}{136mm}
\caption{Orbital parameters of PSR J0514$-$4002}
\label{table:orbital_par1}
\begin{tabular}{l|c|c|c}
\hline
Parameter                                              & \cite{Freire_etal_04} &\cite{Freire_etal_07}       & This work  \\
                                                       & (Period analysis)     & (Coherent timing analysis) &         \\ \hline
Orbital period ($P_b$)                                 & 18.7850(8)            &  18.7851915(4)     &18.7851(1) \\
(days)                                                 &                       &                    &[0.00003]$^\ddagger$   \\\hline
Eccentricity  ($e$)                                    & 0.889(2)              & 0.8879773(3)       & 0.8879(2)\\
                                                       &                       &                    &[0.000005]$^\ddagger$ \\\hline
Longitude of periastron ($\omega$)                     & 82(1)                 & 82.266550(18)      & 82.20(6) \\
(\degr)                                                &                       &                    & [0.002]$^\ddagger$          \\\hline
Semi major axis of the orbit                           & 36.4(2)               & 36.2965(9)         & 36.28(1)   \\
projected along LOS (${a_1 sin(i)}/{c}$)               &                       &                    & \\
(light-seconds)                                        &                       & &\\\hline
Pulsar period ($P_o$)                                  & 4.990576(5)           &4.990575114114(3)   & 4.990575(4) \\
(ms)                                                   &                       && \\\hline
Epoch of periastron passage ($T_o$)                    & 52984.46(2)           & -                  & 52984.5(1)    \\
(MJD)                                                  &                       &                    & [0.02]$^\ddagger$\\ \hline
\end{tabular}

$\dagger$ : The uncertainty quoted in the bracket is on the last significant digit of the concerned parameter.\\
$\ddagger$ : The step size used for comparing the simulation with the observation (Sect. \ref{sec:det_orbital_para}). 
\end{minipage}
\end{table*}
\begin{table*}
\begin{minipage}{136mm}
\caption{Orbital parameters of 47 Tuc T}
\label{table:orbital_par2}
\begin{tabular}{l|c|c|c}
\hline
Parameter                                              & \cite{Freire_etal_b}    & \cite{Freire_etal_a}       & This work  \\
                                                       & (Acceleration analysis) & (Coherent timing analysis) &             \\ \hline
Orbital period ($P_b$)                                 & 1.12(3)              & 1.126176785(5)             & 1.126175(2)   \\
(days)                                                 &                      &                            &[0.0000005]$^\ddagger$ \\\hline
Eccentricity  ($e$)                                    & -                    & 0.00038(2)                 & 0.0000(8)     \\
                                                       &                      &                            &[0.0001]$^\ddagger$\\\hline
Longitude of periastron ($\omega$)                     & -                    &  63(3)                     & 63.0(1)       \\
(\degr)                                                &                      &                            & [0.03]$^\ddagger$ \\\hline
Semi major axis of the orbit                           & 1.33(4)              & 1.33850(1)                 & 1.337(2)    \\
projected along LOS (${a_1 sin(i)}/{c}$)               &                      &                            & \\
(light-seconds)                                        &                      &                            &\\\hline
Pulsar period ($P_o$)                                  & 7.588476(4)          & 7.588479792132(5)          & 7.58848(2) \\
(ms)                                                   &                      &                            & \\\hline
Epoch of periastron passage ($T_o$)                    & 51000.3173(2)        & 51000.317049(2)            & 51000.317(2)   \\
(MJD)                                                  &                      &                            &    [0.0001]$^\ddagger$\\ \hline
\end{tabular}

$\dagger$ : The uncertainty quoted in the bracket is on the last significant digit of the concerned parameter.\\
$\ddagger$ : The step size used for comparing the simulation with the observation (Sect. \ref{sec:det_orbital_para}).
\end{minipage}
\end{table*}
\section{Discussion}    \label{sec:discussion}
The orbital parameters determined in this paper and those 
determined by \cite{Freire_etal_04} and \cite{Freire_etal_07} 
for PSR J0514$-$4002 are listed in Table. \ref{table:orbital_par1}. 
For PSR J0514$-$4002, we have used similar data to those used 
by \cite{Freire_etal_04} (Sect. \ref{sec:det_orbital_P}). 
The orbital parameters determined by us are close to their 
determination within the error quoted by them. But our 
results are more accurate and are close to the values 
obtained by \cite{Freire_etal_07} who have used a much longer 
data stretch from regular observations with the GBT for 
about two years. Table. \ref{table:orbital_par2} compare 
the orbital parameters determined by us with those obtained 
by, \cite{Freire_etal_b} and \cite{Freire_etal_a} for 47 Tuc T. 
Our result agree with \cite{Freire_etal_b}, but are more 
accurate and closer to the values predicted by \cite{Freire_etal_07}, 
who used coherent timing analysis for orbital parameter 
determination. Note that the small eccentricity of 47 Tuc T 
could only be found from the coherent timing solution. Our 
method of orbital parameter determination has the following 
features :

(1) The procedure for determination of binary orbital 
parameters outlined in this paper utilises the measurements 
of $P_{obs}$ at given observing epoch and does not require 
any information about the period derivatives in contrast 
to the method described by \cite{Freire_etal_b}. It may 
at first sight be surprising that period derivatives do 
not help to constrain the final orbital solution. This 
can be understood by examining the accuracy of the 
measurement which is limited by the period variation 
over the length of a single observing session. Clearly, 
the period derivatives implied by the $P_{obs}$ versus 
$\phi$ curves already have smaller errors than this, 
since one is looking at period variations over the $P_b$ 
time scale. However, period derivatives clearly plays a role 
in the work by \cite{Freire_etal_b} in determining orbital 
phases and $P_b$, which in our method comes from the roughness 
search.
 
(2) Unlike the method used by \cite{Freire_etal_b}, which 
works for nearly circular binary orbits, this method works for
binary orbit with any eccentricity. For example, our method 
worked well for the binary orbit with highest known eccentricity 
(PSR J0514$-$4002 with $e\sim$ 0.888), and also for an orbit 
with lower eccentricity (PSR 47 Tuc T with $e\sim$ 0).

(3) The accuracy of the determined orbital parameter values 
are subject to the sampling of the binary orbit. Our method 
works with random sampling of the orbit. A small number of 
data points are required for determination of orbital parameters 
in our method. In case of PSR J0514$-$4002, our method converged 
even for 5 random data points. 

(4) The computation involves only one dimensional searches 
and linear least square fits \footnote{The code we have used 
consists of several stand alone programs in the 'octave' 
(matlab like) language. These programs have not been linked 
to make up a pipeline. Readers interested in the code may 
contact the authors.}.

\section{Acknowledgments}
We thank Yashwant Gupta for the data on PSR J0514$-$4002
used in this paper and for his comments. We thank Paulo C. 
Freire for a very useful discussion during his visit to NCRA. 
We are thankful to Subhashis Roy for critical reading of the 
draft and to Jayanta Roy for helping with coding. We are also 
thankful to an anonymous referee for his suggestions towards 
improvements of the paper. 


\bsp

\appendix
\section[]{Fitting an ellipse to the Even versus Odd data}
\label{sec:appendixA}
For fitting an ellipse to a set of points 
($P_{obs}^{even}$ versus $P_{obs}^{odd}$) we use the 
information that the origin of the ellipse will be at 
(0, $(eGM/h)\:cos\:\omega$), and the major and minor 
axis of the ellipse will be $(GM/h)\:sin\:\omega$ 
and $(GM/h)\:cos\:\omega$. Using this information we 
get an expression which is linear in parameters and 
hence is easy to fit. The ellipse will be of the form,
\begin{equation}  
\frac{X^2}{{(\frac{GM}{h}\:sin\:\omega)}^2} + \frac{{(Y-e\frac{GM}{h}cos\:\omega)}^2}{{{(\frac{GM}{h} \:cos\:\omega)}^2}}=1
\label{eqn12}
\end{equation}
Replacing $(GM/h)\:sin\:\omega=a$, $(GM/h)\:cos\:\omega=b$, $(eGM/h)\:cos\:\omega=d$ we have,
\begin{equation}  
\frac{X^2}{a^2} + \frac {{(Y-d)}^2}{b^2}=1
\label{eqn13}
\end{equation}
Which can easily be simplified to the form,
\begin{equation}  
AX^2+BY^2+CY=1
\label{eqn14}
\end{equation}
Where $A=(1/a^2)/(1-b^2/d^2),B=(1/b^2)/(1-b^2/d^2), 
C=-(2d/b^2)/(1-b^2/d^2)$. We use the singular value 
decomposition method, as described by \cite {Press_etal} 
(\cite{Freire_etal_b} used this method) to determine 
$A$, $B$ and $C$.\footnote {While doing the ellipse 
fitting for the real data we used ${P_{obs}^{odd}}$ 
versus mean subtracted ${P_{obs}^{even}}$ data to 
avoid numerical problems.}
${\mathbf{\chi ^2}}$ in this case is defined as,
\begin{equation}  
\chi ^2=\sum_{i=1}^{N} ((A\:({P_{obs}^{odd}})_i^2+B\:({P_{obs}^{even}})_i^2+C({P_{obs}^{even}})_i)-1)^2
\label{eqn15}
\end{equation}
Here, ${\mathbf{\chi^2}}$ means deviations of the points 
normal to the ellipse. Criterion of minimising the 
${\mathbf{\chi ^2}}$ value gave us satisfactory results. 
From parameters of the fitted ellipse ($A$, $B$ and $C$) 
we determine a, b and c and obtain $e$, $\omega$ values 
as, $e=d/b$ and $\omega=tan^{-1}(a/b)$.

\section[]{Illustration of the straight line nature of observed
pulsar period versus simulated radial velocity plot}
\label{sec:appendixB}
Here we explain the straight line nature of $P_{obs}$ 
versus ${v_{r}}_{s}$ plot and interpret the slope and 
intercept in terms of the orbital parameters. We consider 
the binary orbit of the pulsar, where $m_p$ and $v_{p}$ 
are the mass and velocity of the pulsar and $m_c$ and $v_c$ 
are the same for the companion. $a$ is the semi major axis 
of the pulsar orbit relative to the companion and $a_1$ 
is the semi major axis of the pulsar relative to the center
of mass. Using the standard relation between mass and specific 
angular momentum in a Kepler orbit we make the following 
illustrations for the relative orbit of the pulsar with 
respect to the companion. 
\begin{eqnarray}
\frac{GM}{h}= \frac{G(m_p+m_c)}{\sqrt{a(1-e^2)G(m_p+m_c)}}\\
v_{r}=(v_{p}-v_c)=\frac{m_p+m_c}{m_c} v_{p}\\
a=a_1\frac{m_p+m_c}{m_c}
\label{eqn16}
\end{eqnarray}

Substituting ${GM}/{h}$ (from Eqn. B1) in Eqn. \ref{eqn6},
\begin{equation}  
v_{r}=\sqrt{\frac{G(m_p+m_c)}{a(1-e^2)}} \times {v_{r}}_{s} 
\label{eqn17}
\end{equation}
Therefore velocity of the pulsar $v_{p}$ can be obtained from Eqn. B2
as, 
\begin{equation}  
v_{p}=\frac{m_c}{m_p+m_c}\sqrt{\frac{G(m_p+m_c)}{a(1-e^2)}} \times {v_{r}}_{s} 
\label{eqn18} 
\end{equation}
Projected velocity of the pulsar in the line of sight direction ($v_l$) is given by,
\begin{equation}  
v_{l}=v_{p} \times sin\:i=\frac{m_c}{m_p+m_c}\sqrt{\frac{G(m_p+m_c)}{a(1-e^2)}} \times {v_{r}}_{s} \times sin\:i
\label{eqn19}
\end{equation}
Therefore, $v_{l}$ versus ${v_{r}}_{s}$ is a straight line with slope ($S$),
\begin{equation}  
S=\frac{m_c}{m_p+m_c} \sqrt{\frac {Gm_c}{a_1(1-e^2)}} sin\:i
\label{eqn20}
\end{equation}
So $P_{obs}$ versus ${v_{r}}_{s}$ will also be a straight line with slope ($S_{fit}$),
\begin{equation}  
S_{fit}=\frac{P_{o}}{c}\times S
\label{eqn21}
\end{equation}
But $P_b$ and $a_1$ are related by,
\begin{equation} 
{P_b}^2=\frac{4\pi^2{a_1}^3}{G} \left(\frac {(m_p+m_c)^2}{{m_c}^3}\right)
\label{eqn22} 
\end{equation}
Therefore, from Eqn. \ref{eqn21} and Eqn. \ref{eqn22},
\begin{equation} 
        {(a_1 sin\:i)}^2=\frac{P_b^2 S_{fit}^2(1-e^2)c^2}{4\pi^2 P_{o}^2} 
\label{eqn23} 
\end{equation}
\label{lastpage}       
\end{document}